\documentstyle[aps,preprint]{revtex}

\preprint{hep-th/9706052}

\begin{document}

\title{Liouville-Neumann   Approach   to  the
Nonperturbative   Quantum   Field
Theory\footnote{To appear in the proceedings of APCTP-ICTP Joint
International Conference '97 on {\it Recent Developments in Nonperturbative
Quantum Field Theory}, May 26-30 , 1997, Seoul.}}

\author{Sang Pyo Kim\footnote{E-Mail: sangkim@knusun1.kunsan.ac.kr}}

\address{Department of Physics,
Kunsan National University,\\
Kunsan 573-701, Korea}

\maketitle
\begin{abstract}
We present a nonperturbative field theoretic method based on the
Liouville-Neumann (LN) equation. The LN approach provides
a unified formulation of nonperturbative quantum fields
and also nonequilibrium quantum fields, which makes
use of  mean-field type equations and whose results at the lowest
level are identically the same as those of the Gaussian effective potential
approach and the mean-field approach. The great advantageous point
of this formulation is its readiness of applicability to time-dependent
quantum systems and to finite temperature field theory, and its possibility
to  go beyond the Gaussian approximation.
\end{abstract}

\section{Introduction}

In this conference the terminology of {\it nonperturbative} has
a diverse meaning. In a different community of physics
it has been used in a different sense. Sometimes it means {\it topological},
{\it solitonic}, {\it exactly solvable}, or {\it dual}. However,
in this talk we shall use the term {\it nonperturbative} to imply
that the pertinent propagator represents a resummation of
loop diagrams analogous to the Gaussian effective
potential \cite{stevenson,cea} and the mean-field approaches
\cite{chang}. The Gaussian effective potential
approach has already been reviewed in detail by Prof.
Jae Hyung Yee \cite{yee}.

Our formulation comes from the following observation
that the Liouville-Neumann
(LN) equation enables one to find the exact quantum states
of the Schr\"{o}dinger equation for both time-independent and
time-dependent quantum systems.
We work in the Schr\"{o}dinger-picture
\begin{equation}
i \hbar \frac{\partial}{\partial t} \psi (t) = \hat{H} \psi(t).
\end{equation}
It is known that  any operator  satisfying the following equations
\begin{eqnarray}
 i \hbar \frac{\partial}{\partial t} \hat{O}
+ [\hat{O}, \hat{H}] - i \hbar \frac{\dot{\eta}}{\eta} \hat{O} = 0,
\nonumber\\
\hat{O} \vert \eta(t) \rangle = \eta (t) \vert \eta (t) \rangle
\end{eqnarray}
provides the exact quantum states in terms of its eigenstates.
In particular, when $\dot{\eta} = 0$, the above equation reduces
nothing but to the LN equation
\begin{equation}
 i \hbar \frac{\partial}{\partial t} \hat{O}
+ [\hat{O}, \hat{H}] = 0.
\label{ln eq}
\end{equation}
Furthermore, it is the same equation that the density operators
should satisfy.
All these together seem to suggest that the LN equation
may allow in a unified way a formulation quite useful
not only in quantum mechanics but also in quantum field theory,
irrespective of their time-dependency.

In this talk we develop further the LN formulation, which
was applied to a free quantum field in an expanding FRW
Universe \cite{kim1}, so that it can be applied even to
a self-interacting scalar field in (1+1) dimensions.
The key idea for this extension to nonlinearity
has already been manifested in a quantum Duffing oscillator
\cite{kim2}.

\section{Anharmonic Oscillators as Toy Model}

To exploit the main idea but to keep complexity and computation
minimal, we consider an anharmonic oscillator called Duffing oscillator
as a $(0 + 1)$-dimensional toy model for a self-interacting quantum field
in $(3 +1)$ dimensions. It is just a quantum mechanical system. However,
it exhibits all the important features of a non-trial quantum field.
The Duffing oscillator has already been used as a toy model
to test  various field theoretical perturbation methods
\cite{bender}.
Let us now consider a time-independent quantum Duffing oscillator
\begin{equation}
\hat{H} = \frac{\hat{p}^2}{2m} + \frac{m \omega^2 \hat{q}^2}{2}
+ \frac{m \lambda \hat{q}^4}{4}.
\label{duff}
\end{equation}
Here $m$ is an apparent mass-parameter, $\omega^2$
a  frequency squared, and $\lambda$
a coupling constant. In quantum field  theory in the Minkowski spacetime,
$m =  1$ and $\omega^2$  is the  real mass squared  as a coupling  constant,
whereas in  an expanding  Universe $m  = R(t)$ or $R^3  (t)$, $R(t)$  being the
 scale
factor of the Universe in (1+1) and (3+1) dimensions.

The basic building block both for the Gaussian effective potential
approach and for the LN approach is to find an appropriate
(optimizing) Fock space
\begin{equation}
\hat{a}^{\dagger} = u \hat{p} - m \dot{u} \hat{q},
\hat{a} = u^* \hat{p} - m \dot{u}^* \hat{q}.
\label{cr-an}
\end{equation}
We require the common commutation relation
$ \left[ \hat{a},\hat{a}^{\dagger} \right] =1$, which
leads to the boundary condition
\begin{equation}
\hbar m \bigl( \dot{u}^* u - \dot{u} u^* \bigr) = i.
\end{equation}
Once given a Hamiltonian, one can evaluate the effective potential
\begin{equation}
V_{eff} =  \min_{\vert \psi \rangle}
\langle \psi \vert \hat{H} \vert \psi \rangle
\end{equation}
where the variation is taken with respect to all the normalized state
$\langle \psi \vert \psi \rangle = 1$.
One additional condition one can further impose is either $\langle \psi
\vert \hat{q} \vert \psi \rangle = 0$ or
$\langle \psi \vert \hat{q} \vert \psi \rangle = q_c$. $q_c$
corresponds to a classical background field of the field theory.
We perform the expectation value of the Hamiltonian to find
the effective potential
\begin{equation}
V_{eff} =
\frac{m}{2} \hbar^2 \dot{u}^* \dot{u} +
\frac{m \omega^2}{2} \hbar^2 u^* u + \frac{3 \lambda}{4}
(\hbar^2 u^* u) ^2.
\end{equation}
The problem of finding the Fock space that optimizes
the effective potential can be solved by varying $\Omega$
of trial wave functions
\begin{eqnarray}
u = \frac{1}{\sqrt{2 \hbar m \Omega}} e^{ - i \Omega t}, (\Omega > 0)
\nonumber\\
u = \frac{1}{\sqrt{- 2 \hbar m \Omega}} e^{  i \Omega t}, (\Omega < 0).
\label{sol}
\end{eqnarray}
By minimizing the effective potential, $(\frac{\partial V_{eff}}{\partial
\Omega} = 0)$, we obtain the gap equation \cite{caswell}
\begin{equation}
\Omega^3 - \omega^2 \Omega - \frac{3 \hbar \lambda}{2m} = 0.
\label{gap eq}
\end{equation}
For the positive $\omega^2$, we have a real postive
$\Omega$, and for the negative $\omega^2$, we have always a real negative
$\Omega$ for both a weak and a strong coupling constant $\lambda$,
and a real positive $\Omega$ for a strong coupling constant $\lambda$.

In the LN formulation we wish to find the LN operators that
satisfy Eq. (\ref{ln eq}).
We normal order the Hamiltonian operator and regroup
the quadratic $\hat{H}_2$ , the quartic $\hat{H}_4$
and the constant vacuum expectation value as
\begin{eqnarray}
\hat{H}_2 &=& \Bigl[ m \hbar^2 \dot{u}^* \dot{u}
+ m \omega^2 \hbar^2 u^* u + 3 m \lambda
(\hbar^2 u^* u)^2 \Bigr] \hat{a}^{\dagger} \hat{a}
\nonumber\\
&-& \Bigl[\frac{m}{2} \hbar^2 \dot{u}^{*2}
+ \frac{m \omega^2 }{2} \hbar^2 u^{*2}
+ \frac{3 m \lambda}{2} (\hbar^2 u^* u) \hbar^2 u^{*2}
\Bigr] \hat{a}^{\dagger 2}
\nonumber\\
&-& \Bigl[\frac{m}{2} \hbar^2 \dot{u}^{2}
+ \frac{m \omega^2 }{2} \hbar^2 u^{2}
+ \frac{3 m \lambda}{2} (\hbar^2 u^* u) \hbar^2 u^{2} \Bigr] \hat{a}^{ 2},
\nonumber\\
\hat{H}_4 &=& \frac{m \lambda}{4} \sum_{k = 0}^{4} {4 \choose k} \hbar^4
u^{* (4 - k)} (- u)^k \hat{a}^{\dagger (4 - k)} \hat{a}^k.
\end{eqnarray}
To find the exact LN operators is difficult due to the
infinite group structure of anharmonic oscillators.
So we rely on the approximate LN equation
\begin{equation}
i \hbar \frac{\partial}{\partial t} \hat{a}^{\dagger}
+ \left[\hat{a}^{\dagger}, \hat{H}_2 \right]
= 0,
i \hbar \frac{\partial}{\partial t} \hat{a}
+ \left[\hat{a}, \hat{H}_2 \right]
= 0.
\label{lowest}
\end{equation}
The LN equation (\ref{lowest}) leads to the mean-field equation
\begin{equation}
\ddot{u} + \omega^2 u + 3 \lambda (\hbar^2 u^* u) u = 0.
\label{mean}
\end{equation}
The mean-field equation has a solution (\ref{sol}), which also
satisfies the same gap equation (\ref{gap eq}).
This means that the LN formulation at the lowest order
gives rise to identically the same results as the Gaussian effective
potential and the mean-field approaches.
In fact, we see that the ground state wave function in the Fock space
is given by
\begin{equation}
\Psi_{(0)} (q) = \Bigl( \frac{m \Omega}{ \pi \hbar} \Bigr)^{1/4}
e^{- \frac{m \Omega}{2 \hbar} q^2}.
\end{equation}

The merit of the LN formulation is that it can directly and
straightforward be applied to time-dependent quantum systems, too.
The only change that is to be made
to the problem of a time-dependent quantum
Duffing oscillator, $(m(t), \omega^2 (t), \lambda (t))$,
is the mean-field equation, which now reads
\begin{equation}
\ddot{u} + \frac{\dot{m}}{m} \dot{u} +
\omega^2 u + 3 \lambda (\hbar^2 u^* u) u = 0.
\label{mean}
\end{equation}

A unifying feature of the LN formulation is that one can define
the density operator easily in terms of the creation and annihilation
operators which have already been chosen to satisfy the LN equation:
\begin{equation}
\hat{\rho} = \frac{e^{- \beta \omega_0 \hat{a}^{\dagger} \hat{a}}}{
{\rm Tr} e^{ - \beta \omega_0 \hat{a}^{\dagger} \hat{a}}}.
\end{equation}
In the context of quantum field theory, this implies that
the LN formulation has a close connection with finite
temperature field theory.

As the second case, we consider the quantum state that
describes fluctuations around a classical background.
The variable $q$ is divided into
\begin{equation}
q = q_c + q_f.
\end{equation}
$q_c$ is the classical background and $q_f$ is the quantum fluctuation.
$\hat{q}_c$ and $\hat{q}_f$ commute each other.
Likewise we divide the Hamiltonian into the sum of the classical
background part and the fluctuation part and a perturbation:
\begin{eqnarray}
\hat{H} &=& \Bigl[ \frac{\hat{p}_c^2}{2m} + \frac{m \omega^2}{2}
 \hat{q}_c^2
+ \frac{m \lambda}{4} \hat{q}_c^4 \Bigr]
\nonumber\\
&+& \Bigl[\frac{\hat{p}_f^2}{2m} + \frac{m \omega^2}{2} \hat{q}^2
+ \frac{3m \lambda}{2} \hat{q}_c^2 \hat{q}_f^2
+ \frac{m \lambda}{4} \hat{q}^4 \Bigr]
\nonumber\\
&+& \Bigl[ \frac{\hat{p}_c \hat{p}_f}{m}
+ m \omega^2 \hat{q}_c \hat{q}_f +
m \lambda (\hat{q}_c^3 \hat{q}_f + \hat{q}_c \hat{q}_f^3) \Bigr]
\end{eqnarray}
We take the symmetric quantum state around $q_c$, that is,
\begin{equation}
\langle \hat{q} \rangle = \langle \hat{q}_c \rangle = q_c,
{}~ \langle \hat{q}_f \rangle = 0.
\end{equation}
It also holds that any odd power of $\hat{q}_f$ or $\hat{p}_f$
yields the zero-expectation value. In fact, the quantum state of the
original variable $q$ is a coherent state.
We may quantize both the classical background and the fluctuation
as in the case of symmetric
vacuum state above. Then the change that is to be be made to Eq. (\ref{mean})
is the frequency, which has now a mean-field value
\begin{equation}
\omega_m^2 = \omega^2 + 3 \lambda q_c^2.
\end{equation}
This result can also be obtained in the mean-field approach
by applying the factorization theorem to the original Hamiltonian
(\ref{duff})
\begin{equation}
\hat{q}^4 = 6 \langle \hat{q}^2 \rangle \hat{q}^2 - 8 \langle \hat{q}^3
\rangle \hat{q} + 6 \langle \hat{q} \rangle^4 - 3 \langle \hat{q}^2
\rangle^2,
\end{equation}
and by inserting $\langle \hat{q}^2 \rangle = q_c^2$
and $\langle \hat{q} \rangle = \langle \hat{q}^3 \rangle = 0$.
The time-dependent case can be treated similarly.

\section{$\frac{{\mu}^2}{2} \Phi^2 + \frac{\lambda}{4} \Phi^4$ Field Theory}

To extend the LN formulation to quantum field theory,
we consider a field theoretical model of a (1+1)-dimensional field.
Our model has the Hamiltonian
\begin{equation}
H = \int d x \Biggl[\frac{1}{2} \Pi^2 ( x, t)
+  \frac{m^2}{2} \Phi^2 (x, t) +
\frac{1}{2} ( \nabla \Phi)^2 +
\frac{\lambda}{4} \Phi^4 (x, t) \Biggr].
\end{equation}
The technical idea behind the extension from a quantum system
to a quantum field is that  the Hamiltonian can be rewritten
as a sum of either decoupled or coupled anharmonic oscillators in terms of
Fourier-modes. To see how this
work we consider separately a free scalar field $(\lambda = 0)$
and a self-interacting scalar field $(\lambda \neq 0)$.

\subsection{Free Scalar Field}

We use the box normalization of the field confined
into a square well of length $L$. We denote the orthonormal basis
by
\begin{equation}
\xi_{\alpha} = \sqrt{\frac{2}{L}} \cos \Bigl( \frac{(2n -1) \pi x}{L} \Bigr)
{}~or~~ \sqrt{\frac{2}{L}} \sin \Bigl(\frac{2n \pi x}{L} \Bigr),
\end{equation}
where $\alpha = \frac{n \pi}{L}$ and $\sum_{\alpha} = \sum_{n = 1}^{\infty}$.
We decompose the field into modes
\begin{equation}
\Phi ( x, t) = \sum_{\alpha} \phi_{\alpha} (t) \xi_{\alpha} (x).
\end{equation}
The Hamiltonian is decomposed into a collection of harmonic
oscillators
\begin{equation}
H = \sum_{\alpha} H_{\alpha} = \sum_{\alpha}
\frac{1}{2}\pi_{\alpha}^2 + \frac{m^2 + {\alpha}^2}{2} \phi_{\alpha}^2.
\end{equation}
We quantize the system according to the Schr\"{o}dinger-picture
\begin{equation}
i \hbar \frac{\partial}{\partial t} \vert \Psi \rangle =
\hat{H} \vert \Psi \rangle.
\end{equation}
The wave function as expected is easily found to be
$\vert \Psi \rangle = \prod_{\alpha} \vert \psi_{\alpha} \rangle$
whose wave function for each mode obeys the Schr\"{o}dinger equation
with the harmonic oscillator Hamiltonian $H_{\alpha}$.
Each position operator has a Fock space representation
\begin{equation}
\hat{\phi}_{\alpha} = (- i \hbar) \Bigl( u^*_{\alpha}
\hat{a}^{\dagger}_{\alpha}
- u_{\alpha} \hat{a}_{\alpha} \Bigr).
\end{equation}

Firstly, we assume  $\langle \hat{\phi}_{\alpha} \rangle = 0$, which
in turn implies $\langle \hat{\Phi} \rangle = 0$.
The nontrivial contributions come only from even power of field operators;
for instance, the correlation function is given by
\begin{equation}
\langle \hat{\Phi}^2 \rangle = \hbar^2 \sum_{\alpha}
u_{\alpha}^* u_{\alpha}
= \hbar^2 \sum_{n = 1}^{\infty} u^*_{\frac{n \pi}{L}} u_{\frac{n \pi}{L}}.
\end{equation}
We calculate the expectation value of the Hamiltonian with respect to
the vacuum state $\vert 0, t \rangle = \prod_{\alpha}
\vert 0_{\alpha}, t \rangle$ to obtain the effective potential
\begin{equation}
V_{eff} = \frac{\hbar}{2} \sum_{n = 1}^{\infty} \Biggl(
m^2 + \Bigl(\frac{n\pi}{L}\Bigr)^2 \Biggr)^{1/2}.
\end{equation}
It is the sum of the vacuum fluctuations of each oscillator,
which requires a renormalization.

Secondly, we consider the case of $ \langle \hat{\phi}_{\alpha} \rangle
= \phi_c$, i.e. $\langle \hat{\Phi} \rangle = \phi_c$,
we divide the field into $\phi_{\alpha} = \phi_c + \phi_{f \alpha}$.
After quantizing the fluctuations, we obtain the effective Hamiltonian
\begin{equation}
V_{eff} = \int dx \Biggl( \frac{1}{2} \pi_c^2 + \frac{m^2}{2}
\phi_c^2 \Biggr) +
\frac{\hbar}{2} \sum_{n = 1}^{\infty} \Biggl(
m^2 + \Bigl(\frac{n\pi}{L}\Bigr)^2 \Biggr)^{1/2}.
\end{equation}

\subsection{Self-Interacting Scalar Field}

Firstly, we consider the case $\langle \hat{\Phi} \rangle = 0$.
Recollecting that only even power of field operators contribute,
we compute
\begin{equation}
\langle \hat{\Phi}^4 \rangle = 3 \Biggl[ \sum_{\alpha}
\hbar^2 u^*_{\alpha} u_{\alpha} \Biggr]^2
\end{equation}
The effective potential now reads that
\begin{equation}
V_{eff} =  \frac{\hbar^2}{2} \sum_{\alpha} \Biggl[
\dot{u}^*_{\alpha} \dot{u}_{\alpha}
+ \Bigl( m^2 + {\alpha}^2 \Bigr) u^*_{\alpha} u_{\alpha} \Biggr]
+ \frac{3 \hbar^4 \lambda}{4}  \Biggl [\sum_{\alpha}
u^*_{\alpha} u_{\alpha} \Biggr]^2.
\end{equation}

Either from the minimization of the effective potential or from
the LN equation at the quadratic level (lowest order),
we obtain the field equation
\begin{equation}
\ddot{u}_{\alpha} + \Bigl[m^2 + \Bigl(\frac{n \pi}{L} \Bigr)^2
\Bigr] u_{\alpha} + 3 \hbar^2 \lambda I_0 u_{\alpha} = 0,
\end{equation}
where
\begin{equation}
I_0 = \sum_{\alpha}
 u_{\alpha}^* u_{\alpha}
=  \sum_{n = 1}^{\infty} u^*_{\frac{n \pi}{L}} u_{\frac{n \pi}{L}}.
\end{equation}
Since $I_0$ contains an infinite quantity, it requires the renormalization
of coupling constants $m^2$ and $\lambda$. Then the renormalized field
equation should read
\begin{equation}
\ddot{u}_{\alpha} + \Bigl[m^2_R + \Bigl(\frac{n \pi}{L} \Bigr)^2
\Bigr] u_{\alpha} + 3 \hbar^2 \lambda_R I_{0,R} u_{\alpha} = 0.
\end{equation}
We did not show the renormalization procedure in detail.

Secondly, in the case $\langle \hat{\Phi} \rangle = \phi_c$, we repeat
the same procedure as for the anharmonic oscillator but
keep in mind that we now treat a field rather than a finite
degrees of freedom.
We divide the field into a classical background field and a fluctuation
field
\begin{equation}
\Phi ({\bf x}, t) = \phi_c (t) + \Phi_f ({\bf x}, t).
\end{equation}
$\langle \hat{\Phi} \rangle = \phi_c$
and $\langle \hat{\Phi}_f \rangle = 0$ means a condensation of bosonic
particles and a (collective) coherent motion of $\phi_c$.
We rewrite the Hamilonian density as
\begin{eqnarray}
\hat{\cal H} &=& \Bigl[ \frac{1}{2} \hat{\pi}_c^2 + \frac{m^2}{2}
 \hat{\phi}_c^2
+ \frac{ \lambda}{4} \hat{\phi}_c^4\Bigr]
\nonumber\\
&+& \Bigl[ \frac{1}{2} \hat{\Pi}_f^2 + \frac{m^2}{2} \hat{\Phi}^2
+ \frac{1}{2} (\nabla \hat{\Phi}_f)^2
+ \frac{3\lambda}{2} \hat{\phi}_c^2 \hat{\Phi}_f^2
+ \frac{ \lambda}{4} \hat{\Phi}_f^4 \Bigr]
\nonumber\\
&+& \Bigl[ \hat{\pi}_c \hat{\Phi}_f
+  m^2 \hat{\phi}_c \hat{\Phi}_f +
 \lambda (\hat{\phi}_c^3 \hat{\Phi}_f
+ \hat{\phi}_c \hat{\Phi}_f^3) \Bigr].
\end{eqnarray}
By quantizing the fluctuation field, we obtain the effective potential
of fluctuation
\begin{eqnarray}
V_{eff} =  \frac{\hbar^2}{2} \sum_{\alpha} \Biggl[
\dot{u}^*_{\alpha} \dot{u}_{\alpha}
+ \Bigl( m^2 + {\alpha}^2  + 3 \lambda \phi_c^2 \Bigr)
u^*_{\alpha} u_{\alpha} \Biggr]
\nonumber\\
+ \frac{3 \hbar^4 \lambda}{4} \Biggl[\sum_{\alpha}
u^*_{\alpha} u_{\alpha} \Biggr]^2.
\end{eqnarray}
It should be noted that the effective mass squared has been changed
to $m^2 + 3 \lambda \phi_c^2$. We find the renormalized
field equation
\begin{equation}
\ddot{u}_{\alpha} + \bigl[m^2_R + \alpha^2 + 3 \lambda_R \phi_c^2
\bigr] u_{\alpha}
+ 3 \hbar^2 \lambda_R I_{0, R} u_{\alpha} = 0.
\end{equation}

\subsection{Cosmological Model}

As a time-dependent quantum system, we consider a self-interacting
scalar field in an expanding FRW universe in (1+1) dimensions
\begin{equation}
ds^2 = - dt^2 + R^2 (t) dx^2.
\end{equation}
$R(t)$ is the scale factor of the Universe.
The Hamiltonian takes the form
\begin{equation}
H = \int dx \Biggl[\frac{1}{2R} \Pi^2 ( x, t)
+  \frac{m^2 R}{2} \Phi^2 (x, t) + \frac{1}{2R}
(\frac{d}{dx} \Phi)^2 +
\frac{\lambda R}{4} \Phi^4 (x, t) \Biggr].
\end{equation}
In (3+1) dimensional cosmology, $\Phi$ plays the role of both
an inflaton (a classical background field) and a fluctuation.
As in the Minkowski spacetime, we again divide the field  into
$\Phi ({\bf x}, t) = \phi_c (t) + \Phi_f ({\bf x}, t)$,
where $\langle \hat{\Phi} \rangle = \phi_c$
and $\langle \hat{\Phi}_f \rangle = 0$,
and  obtain the effective potential of fluctuation
\begin{eqnarray}
V_{eff} =
 \frac{\hbar^2 R}{2} \sum_{\alpha} \Biggl[
\dot{u}^*_{\alpha} \dot{u}_{\alpha}
+  \Bigl( m^2 + \frac{{\alpha}^2}{R^2}  + 3 \lambda \phi_c^2 \Bigr)
u^*_{\alpha} u_{\alpha} \Biggr]
\nonumber\\
+ \frac{3 \hbar^4 \lambda R}{4} \Biggl[\sum_{\alpha}
u^*_{\alpha} u_{\alpha} \Biggr]^2.
\end{eqnarray}
It should be noted that the effective mass squared
again has been changed to $m^2 + 3 \lambda \phi_c^2$.
This formulation can be used to treat some cosmological issues
such as preheating mechanism.

\section{Beyond Gaussian Approximation}

We suggest how to go beyond the Gaussian approximation.
We return to the toy model and
construct the LN operators in a perturbative way
\begin{eqnarray}
\hat{A}^{\dagger} = \hat{a}^{\dagger} + \sum_{n = 1}^{\infty}
 \bigl( m \lambda \bigr)^n
\hat{B}_{2n +1}^{\dagger},~
\hat{A} = \hat{a}  + \sum_{n = 1}^{\infty}
 \bigl( m \lambda \bigr)^n
\hat{B}_{2n +1}
\nonumber\\
\hat{B}_{2n +1} = \sum_{k = 0}^{2n +1} b^{(2n+1)}_{k}
\hat{a}^{\dagger (2n +1 - k)} \hat{a}^k.
\end{eqnarray}

The $\hat{U}_{(q+ p = 2)} = \hat{a}^{p\dagger} \hat{q}^q$
basis of the group SU(1,1) leads to Eq. (\ref{lowest}).
In $\hat{U}_{(p+q = 3)}$, we obtain
\begin{equation}
\frac{\partial}{\partial t} \hat{B}_3 = i \Bigl(
\left[ \hat{B}_3, \hat{H}_2 \right]
+  \left[ \hat{a}, \hat{H}_4 \right]
+ m \lambda \left[ \hat{B}_3, \hat{H}_4 \right]_{(3)} \Bigr).
\end{equation}
This leads to an inhomogeneous equation of the form
\begin{equation}
\frac{d}{dt} \vec{B} (t) = i m \lambda {\bf M} (u, u^*) \vec{B} (t)
+ i \vec{D},
\label{inh eq}
\end{equation}
where $\vec{B}$ is a vector of $b_k^3 $, ${\bf M}$ and $\vec{D}$ are a
matrix and a vector depending on $u$ and $u^*$, respectively. We just sketch
the procedure. The first step is to solve
Eq. (\ref{inh eq}) to find the spectrum generating operators
that involve the creation and annihilation operators up to cubic terms.
At the next step we use the holomorphic (Bargmann) representation of the Fock
space
\cite{faddeev}
\begin{equation}
\hat{a}^{\dagger} \rightarrow a^*, \hat{a} \rightarrow
\frac{\partial}{\partial a^*}.
\end{equation}
With respect to the inner product on the space of analytic functions of $a^*$
\begin{equation}
\langle \Psi_1 \vert \Psi_2 \rangle = \int
\frac{da^* da}{2 \pi i} \Psi_1^* (a^*) \Psi_2 (a^*) e^{- a^* a},
\end{equation}
the number states of the Fock space are represented as
$\langle a^* \vert n, t \rangle = \frac{a^{*n}}{\sqrt{n!}}$.
Now at the first level, we may define the nonperturbative ground state by
\begin{equation}
\hat{A} \Psi_0^{(3)} = \Biggl[
\frac{\partial}{\partial a^*} + \sum_{k = 0}^{3}
b_k^{(3)} a^{* (3 - k)} \Bigl(\frac{\partial}{\partial a^*} \Bigr)^k
\Biggr] \Psi^{(3)}_0 = 0.
\end{equation}

Finally, the density operator defined as
\begin{equation}
\hat{\rho} = \frac{e^{ - \beta \Omega_0 \hat{A}^{\dagger} \hat{A}}}{
{\rm Tr} e^{ - \beta \Omega_0 \hat{A}^{\dagger} \hat{A}}}
\label{den op}
\end{equation}
satisfies the LN equation.
It should be noted that Eq. (\ref{den op}) really
goes beyond the quadratic order for the Gaussian-type.

\section*{Acknowledgments}
The author is deeply indebted to Prof. S. K. Kim, Prof. K.-S. Soh
and Prof. J. H. Yee
for many valuable discussions. This work was supported in part
by the Korea Science and Engineering Foundation under
Grant No. 951-0207-056-2 and
by the Non-Directed Research Fund, Korea Research Foundation, 1996.

\end{document}